\documentclass[10pt]{iopart}

\usepackage{graphicx}

\begin{document}

\title[Continuous variable entanglement by radiation pressure]
{Continuous variable entanglement by radiation pressure}

\author{Stefano Pirandola,
Stefano Mancini
\footnote[3]{To
whom correspondence should be addressed (stefano.mancini@unicam.it)},
David Vitali,
and Paolo Tombesi 
}

\address{INFM, Dipartimento di Fisica,
Universit\`a di Camerino,
I-62032 Camerino, Italy
}

\begin{abstract}
We show that the radiation pressure of an intense optical field
impinging on a perfectly reflecting vibrating
mirror is able to entangle in a robust way 
the first two optical sideband modes. Under appropriate conditions,
the generated entangled state is of EPR type 
[A. Einstein, {\it et al.}, Phys. Rev. {\bf 47}, 777 (1935)].
\end{abstract}



\maketitle

\section{Introduction}

The radiation pressure acting on a movable mirror realizes
an optomechanical coupling between the incident optical modes
and the various vibrational modes of the mirror.
The use of this coupling has been proposed many years ago
for the implementation of quantum limited 
measurements of mechanical forces \cite{BKbook}, as
the interferometric detection of gravitational-waves \cite{ABR}, 
or atomic force microscopy \cite{AFM}.
Then optomechanical coupling has been proposed for quantum state
engineering: for example, it has been shown 
that it may lead to
nonclassical states of both the radiation 
field \cite{FAB,PRA94}, 
and the motional degree of freedom of the mirror 
\cite{PRA97}.

The appearance of quantum effects in ponderomotive systems,
paves the way for using them also for quantum information purposes
\cite{QI}. In particular, quantum information can be encoded
in the continuous quadratures
of electromagnetic modes \cite{CVbook}, 
and many quantum communication protocols
can be implemented using entangled states
of optical fields \cite{PVL}. A typical example of continuous
variable entanglement is represented by the 
two-mode squeezed states 
at the output of a nondegenerate parametric amplifier \cite{QO94}. 
However, it has been showed in Refs.~\cite{GMT,ALE,SILVIA}
that radiation pressure can also be used to entangle
two or more modes of a cavity with a movable and perfectly reflecting mirror. 
In the present paper, we get rid of the cavity and we show that
radiation pressure provides a new way
to entangle {\it travelling} electromagnetic modes.
We shall consider the very simple case of an intense optical mode, incident
on a single, vibrating, and perfectly reflecting mirror. 
A vibrational mode of the mirror will induce sideband modes on the
reflected field, which will
be shown to be entangled. Entanglement proves to be very robust with
respect to the thermal noise acting on the mirror. In particular, for a well 
defined interaction time, i.e., a given time duration of the incident field,
the two first sideband modes become an EPR-correlated pair, identical
to the two-mode squeezed state, with the ratio between the optical
and the mechanical frequency playing the role of the squeezing parameter.

The paper is organized as follows.
In Section II we shall derive the effective Hamiltonian of the system.
In Section III we shall exactly solve the dynamics and provide a complete
characterization of the reduced state of the system composed by the two
reflected sideband modes. Section IV is for concluding remarks.

\section{Effective Hamiltonian of the system}

\begin{figure}
\begin{center}
\includegraphics[width=0.5\textwidth]{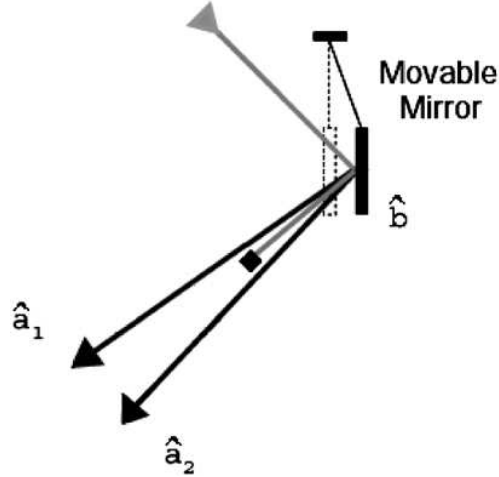}
\end{center}
\vspace{-0.5cm} \caption{\label{fig1} 
Schematic description of the system. A laser field at frequency
$\omega_{0}$ impinges on the mirror oscillating at frequency $\Omega$.
In the reflected field two sideband modes are excited at
frequencies $\omega_{1}=\omega_{0}-\Omega$ and
$\omega_{2}=\omega_{0}+\Omega$.
}
\end{figure}

Let us consider a perfectly reflecting mirror and an intense laser beam
impinging on its surface (see Fig.~\ref{fig1}).
For simplicity we consider only the motion and the elastic deformations
of the mirror taking place along the spatial direction $x$,
orthogonal to its reflecting surface.

The electromagnetic field exerts a force on the mirror 
proportional to its intensity and, at the same time,
it is phase-shifted by
the mirror displacement from the equilibrium position \cite{SAM95,LAW95}.
In the limit of small mirror displacements, and in the interaction
picture with respect to the free Hamiltonian of the electromagnetic field
and the mirror displacement field ${\hat x({\bf r},t)}$
(${\bf r}$ is the coordinate on the mirror surface), one has the
following Hamiltonian
 \cite{PIN99}
\begin{eqnarray}
    {\hat H}&=&-\int\,d^{2}{\bf r}\,
    {\hat P}({\bf r},t){\hat x}({\bf r},t)\,,
    \label{eq:Hini}
\end{eqnarray}
where ${\hat P({\bf r},t)}$ is the radiation pressure force \cite{SAM95}.
All the continuum of electromagnetic modes
with positive longitudinal wave vector $q$ and transverse
wave vector ${\bf k}$
contributes to the radiation pressure force.
Following Ref.\cite{SAM95}, and considering linearly polarized radiation
with the electric field parallel to the mirror surface, we have 
\begin{eqnarray}
        {\hat P}({\bf r},t)&=&-\frac{\hbar}{8\pi^{3}}
        \int d{\bf k}\int dq\int d{\bf k'}\int dq'
        \frac{c^{2}qq'}{\sqrt{\omega\omega'}} 
        ({\bf u}_{k}\cdot{\bf u}_{k'}){\bf u}_{q}
        \nonumber\\
        &\times&
        \left\{{\hat a}({\bf k},q){\hat a}({\bf k}',q')
        \exp[-i(\omega+\omega')t
        +i({\bf k}+{\bf k}')\cdot{\bf r}]\right.
        \nonumber\\
        &&\left.
        +{\hat a}^{\dag}({\bf k},q){\hat a}^{\dag}({\bf k}',q')
        \exp[i(\omega+\omega')t
        -i({\bf k}+{\bf k}')\cdot{\bf r}]\right.
        \nonumber\\
        &&\left. 
        +{\hat a}({\bf k},q){\hat a}^{\dag}({\bf k}',q')
        \exp[-i(\omega-\omega')t
        +i({\bf k}-{\bf k}')\cdot{\bf r}]\right.
        \nonumber\\
        &&\left.
        +{\hat a}^{\dag}({\bf k},q){\hat a}({\bf k}',q')
        \exp[i(\omega-\omega')t
        -i({\bf k}-{\bf k}')\cdot{\bf r}]
        \right\}\,,
        \label{eq:P}
\end{eqnarray}
where ${\hat a}({\bf k},q)$ are continuous mode destruction 
operators obeying the commutation 
relations
\begin{equation}
\left[{\hat a}({\bf k},q),{\hat a}^{\dagger}({\bf k}',q')\right]
=\delta({\bf k}-{\bf k}')\delta(q-q')\,.
\end{equation}
Furthermore, the electromagnetic wave frequencies 
$\omega$ and $\omega'$ are given by $\omega=\sqrt{c^{2}(k^{2}+q^{2})}$
and $\omega'=\sqrt{c^{2}(k'^{2}+q'^{2})}$
($c$ is the light speed in vacuum), 
and ${\bf u}_{k}$, ${\bf u}_{q}$ denote dimensionless
unit vectors parallel to ${\bf k}$, $q$ respectively.

The mirror displacement ${\hat x({\bf r},t)}$ is generally given by a 
superposition of many acoustic modes \cite{PIN99};
however, a single vibrational mode description can be adopted whenever 
detection is limited to a frequency bandwidth
including a single mechanical resonance. 
In particular, focused light beams are able to excite 
Gaussian acoustic modes, in which only a small portion of the mirror,
localized at its center, vibrates. These modes have a small  
waist $w$, a large mechanical quality 
factor $Q$, a small effective mass $M$ \cite{PIN99}, and 
the simplest choice is to choose the fundamental Gaussian mode with 
frequency $\Omega$ and annihilation operator ${\hat b}$, with
$\left[{\hat b},{\hat b}^{\dagger}\right]=1$, 
\begin{equation}
    {\hat x}({\bf r},t)=\sqrt{\frac{\hbar}{2M\Omega}}
    \left[{\hat b} e^{-i\Omega t}+{\hat b}^{\dag}e^{i\Omega t}\right]
    \exp(-r^{2}/w^{2})\,.
    \label{eq:x}
\end{equation}
By inserting Eqs.~(\ref{eq:P}) and (\ref{eq:x}) in Eq.~(\ref{eq:Hini})
and integrating over the variable ${\bf r}$, one obtains 
\begin{eqnarray}
        {\hat H}&=&-\frac{\hbar 
        w^{2}}{8\pi^{2}}\sqrt{\frac{\hbar}{2M\Omega}}
        \int d{\bf k}\int dq\int d{\bf k'}\int dq'
        \frac{c^{2}qq'}{\sqrt{\omega\omega'}} 
        ({\bf u}_{k}\cdot{\bf u}_{k'}){\bf u}_{q}
        \nonumber\\
        &\times&
        \left\{{\hat a}({\bf k},q){\hat a}({\bf k}',q')
        \exp[-i(\omega+\omega')t
        -({\bf k}+{\bf k}')^{2}w^{2}/4]\right.
        \nonumber\\
        &&\left.
        +{\hat a}^{\dag}({\bf k},q){\hat a}^{\dag}({\bf k}',q')
        \exp[i(\omega+\omega')t
        -({\bf k}+{\bf k}')^{2}w^{2}/4]\right.
        \nonumber\\
        &&\left. 
        +{\hat a}({\bf k},q){\hat a}^{\dag}({\bf k}',q')
        \exp[-i(\omega-\omega')t
        -({\bf k}-{\bf k}')^{2}w^{2}/4]\right.
        \nonumber\\
        &&\left.
        +{\hat a}^{\dag}({\bf k},q){\hat a}({\bf k}',q')
        \exp[i(\omega-\omega')t
        -({\bf k}-{\bf k}')^{2}w^{2}/4]
        \right\}
        \nonumber\\
        &&\times\left\{{\hat b}e^{-i\Omega t}
        +{\hat b}^{\dag}e^{i\Omega t}\right\}\,.
        \label{eq:Hint1}
\end{eqnarray}
In common situations, the acoustical waist $w$ is much larger than typical 
optical wavelengths \cite{PIN99}, and therefore we can approximate
$\exp\left\{-({\bf k}\pm {\bf k'})^{2}w^{2}/4\right\}w^{2}/4\pi \simeq 
\delta({\bf k}\pm {\bf k'})$ and then integrate Eq.~(\ref{eq:Hint1})
over ${\bf k'}$, obtaining
\begin{eqnarray}
        {\hat H}&=&-\frac{\hbar}{2\pi}\sqrt{\frac{\hbar}{2M\Omega}}
        \int d{\bf k}\int dq\int dq'
        \frac{c^{2}qq'}{\sqrt{\omega\omega'}} 
        \nonumber\\
        &\times&
        \left\{-{\hat a}({\bf k},q){\hat a}(-{\bf 
        k},q')\exp[-i(\omega+\omega')t]
        \right.
        \nonumber\\
        &&\left.
        -{\hat a}^{\dag}({\bf k},q){\hat a}^{\dag}(-{\bf 
        k},q')\exp[i(\omega+\omega')t]\right.
        \nonumber\\
        &&\left. 
        +{\hat a}({\bf k},q){\hat a}^{\dag}({\bf 
        k},q')\exp[-i(\omega-\omega')t]\right.
        \nonumber\\
        &&\left.
        +{\hat a}^{\dag}({\bf k},q){\hat a}({\bf k},q')
        \exp[i(\omega-\omega')t]
        \right\}\times\left\{{\hat b}e^{-i\Omega t}+{\hat 
        b}^{\dag}e^{i\Omega t}\right\}\,.
        \label{eq:Hint2}
\end{eqnarray}
We now make the Rotating Wave Approximation (RWA), that is, we 
neglect all the terms oscillating in time faster than the mechanical 
frequency $\Omega$. This means averaging the Hamiltonian over a time 
$\tau$ such 
that $\Omega \tau \gg 1$, yielding the following replacements in 
Eq.~(\ref{eq:Hint2})
\begin{equation} \label{rwa}
\exp\left\{\pm i(\omega' \pm \omega \pm \Omega)t\right\} \rightarrow 
\frac{2\pi}{\tau}\delta(\omega' \pm \omega \pm \Omega).
\end{equation}
The parameter $\tau$ is not arbitrary, but its inverse, $1/\tau = 
\Delta\nu_{det}$, is the 
detection bandwidth, that is, the spectral resolution of the 
detection apparatus employed. 

Since $\omega$ and $\omega'$ are positive and $\Omega$ is much 
smaller than typical optical frequencies, the two terms 
$\delta(\omega' + \omega \pm \Omega)$ give no contribution, while the 
other two terms can be rewritten as
\begin{equation}
\frac{2\pi}{\tau}\delta(\omega' - \omega \pm \Omega) =
2\pi\Delta\nu_{det}\delta(q' - 
\bar{q}_{\pm})\frac{\omega'(\bar{q}_{\pm})}{c^{2}\bar{q}_{\pm}},
\end{equation}
where $\bar{q}_{\pm}=\sqrt{(\omega \pm \Omega)^{2}/c^{2}-k^{2}}$.
Integrating over $q'$ we get
\begin{eqnarray}
        {\hat H}&=&-\hbar\Delta\nu_{det}\sqrt{\frac{\hbar}{2M\Omega}}
        \int d{\bf k}\int dq
        \frac{q}{\sqrt{\omega}} 
        \left\{{\hat a}({\bf k},q){\hat a}^{\dag}\left({\bf 
        k},\bar{q}_{+}\right){\hat b}\;
        \sqrt{\omega+\Omega}\right.
        \nonumber\\
        &&\left.+{\hat a}({\bf k},q){\hat a}^{\dag}\left({\bf k},
        \bar{q}_{-}\right){\hat b}^{\dag}\sqrt{\omega-\Omega}
        \right.
        \nonumber\\
        &&\left. 
        +{\hat a}^{\dag}({\bf k},q){\hat a}\left({\bf k},
        \bar{q}_{+}\right){\hat b}^{\dag}
        \sqrt{\omega+\Omega}\right.
        \nonumber\\
        &&\left.
        +{\hat a}^{\dag}({\bf k},q){\hat a}\left({\bf k},
        \bar{q}_{-}\right){\hat b}\sqrt{\omega-\Omega}
        \right\}\,,
        \label{eq:Hint3}
\end{eqnarray}
where we have used the fact that $\omega'(\bar{q}_{\pm})=\omega \pm 
\Omega$.

We now consider the situation where the radiation field incident on 
the mirror is characterized by an intense, quasi-monochromatic, 
laser field with transversal 
wave vector ${\bf k_{0}}$, longitudinal wave vector $q_{0}$, 
cross-sectional area $A$, and power ${\wp}$. Since this component is 
very intense, it can be 
treated as classical and one can approximate
${\hat a}({\bf k},q) \simeq  \alpha({\bf k},q)$ in Eq.~(\ref{eq:Hint3}), 
where (with an appropriate choice of phases)
\begin{equation}
\alpha({\bf k},q) = -i\sqrt{\frac{(2\pi)^{3}{\wp}}{\hbar \omega_{0}cA}}
\delta({\bf k}-{\bf k_{0}})\delta(q-q_{0})\,,
\label{intenso}
\end{equation}
with $\omega_{0}=c\sqrt{{\bf k_{0}}^{2}+q_{0}^{2}}$.

Due to the Dirac delta, the only nonvanishing terms in the 
optomechanical interaction driven by the intense laser beam
involve only two back-scattered waves, that is, the sidebands of the driving 
beam at frequencies 
$\omega_{0}\pm \Omega$, as described by 
\begin{eqnarray}
        {\hat H}&=& i\hbar\Delta\nu_{det}\sqrt{\frac{\hbar}{2M\Omega}}
        q_{0} \sqrt{\frac{{\wp}}{\hbar \omega_{0}cA}}
        \nonumber\\
        &\times&\left\{\sqrt{\frac{\omega_{0}+\Omega}{\omega_{0}}}
        {\hat a}^{\dag}\left({\bf 
        k_{0}},\bar{q}_{+}\right){\hat b}
        +\sqrt{\frac{\omega_{0}-\Omega}{\omega_{0}}}
        {\hat a}^{\dag}\left({\bf 
        k_{0}},\bar{q}_{-}\right){\hat b}^{\dagger}
        \right.
        \nonumber\\
        &&\left. 
        -\sqrt{\frac{\omega_{0}+\Omega}{\omega_{0}}}
        {\hat a}\left({\bf 
        k_{0}},\bar{q}_{+}\right){\hat b}^{\dagger}
        -\sqrt{\frac{\omega_{0}-\Omega}{\omega_{0}}}
        {\hat a}\left({\bf 
        k_{0}},\bar{q}_{-}\right){\hat b}\right\},
        \label{eq:Heff0}
\end{eqnarray}
where now $\bar{q}_{\pm}=\sqrt{(\omega_{0} \pm \Omega)^{2}/c^{2}-k_{0}^{2}}$.
The physical process described by this interaction Hamiltonian is 
very similar to a stimulated Brillouin scattering \cite{PER84}, even though in 
this case the Stokes and anti-Stokes component are back-scattered by 
the acoustic waves at 
reflection, and the optomechanical coupling is provided by the 
radiation pressure
and not by the dielectric properties of the mirror.

In practice, either the driving laser beam and the back-scattered modes
are never monochromatic, but have a nonzero bandwidth. In general the 
bandwidth of the back-scattered modes is determined by the bandwidth 
of the driving laser beam and that of the acoustic mode. However, due 
to its high mechanical quality factor, the spectral width of the 
mechanical resonance is negligible (about $1$ Hz) and, in practice, the 
bandwidth of the two sideband modes $\Delta \nu_{mode}$
coincides with that of the incident laser beam.
It is then convenient to consider this nonzero bandwidth to redefine
the bosonic operators of the Stokes and anti-Stokes modes 
to make them dimensionless, 
\begin{eqnarray}
{\hat a}_{1}&=& 2\pi \sqrt{\frac{2\pi \Delta \nu_{mode}}{cA}}
{\hat a}\left({\bf k_{0}},\bar{q}_{-}\right) =
2 \pi \sqrt{\frac{\Delta q}{A}}{\hat a}\left({\bf k_{0}},\bar{q}_{-}\right)\\
{\hat a}_{2}&=& 2\pi \sqrt{\frac{2\pi \Delta \nu_{mode}}{cA}}
{\hat a}\left({\bf k_{0}},\bar{q}_{+}\right)=
2 \pi \sqrt{\frac{\Delta q}{A}}{\hat a}\left({\bf k_{0}},\bar{q}_{+}\right), 
\end{eqnarray}
so that  Eq.(\ref{eq:Heff0}) reduces to an effective 
Hamiltonian
\begin{equation}
    {\hat H}_{eff}=-i\hbar \chi
    ({\hat a}_{1}{\hat b}-{\hat a}^{\dag}_{1}{\hat b}^{\dag})
    -i\hbar\theta({\hat a}_{2}{\hat b}^{\dag}-{\hat a}^{\dag}_{2}
    {\hat b})\,,
    \label{eq:Heff}
\end{equation}
where the couplings $\chi$ and $\theta$ are given by
\begin{eqnarray}
\label{chi}
\chi &=& q_{0}\Delta\nu_{det}\sqrt{\frac{\hbar}{2M\Omega}}
\sqrt{\frac{{\wp} }{ \Delta \nu_{mode}\hbar \omega_{0}}}
\sqrt{\frac{\omega_{0}-\Omega}{\omega_{0}}}
\nonumber\\
&=&\cos\phi_{0}\sqrt{\frac{{\wp}\Delta\nu_{det}^{2}(\omega_{0}-\Omega)}
{2M\Omega 
c^{2}\Delta\nu_{mode}}} \\
\theta &=& \chi \sqrt{\frac{\omega_{0}+\Omega}{\omega_{0}-\Omega}},
\end{eqnarray}
with $\phi_{0}=\arccos(cq_{0}/\omega_{0})$, 
is the angle of incidence of the driving beam. 
It is possible to verify that with the above definitions, the Stokes 
and anti-Stokes annihilation operators $a_{1}$ and $a_{2}$ satisfy the 
usual commutation relations 
$\left[a_{i},a_{j}^{\dagger}\right]=\delta_{i,j}$.

\section{System dynamics}

Eq.~(\ref{eq:Heff}) contains two interaction terms: the first one,
between modes ${\hat a}_{1}$ and ${\hat b}$,
is a parametric-type interaction
leading to squeezing in phase space \cite{QO94}, and it is
able to generate the EPR-like
entangled state which has been used in the continuous variable teleportation
experiment of Ref.~\cite{FUR98}. The
second interaction term, between modes ${\hat a}_{2}$ and ${\hat b}$,
is a beam-splitter-type
interaction \cite{QO94}, which may degrade the entanglement between
modes ${\hat a}_{1}$ and ${\hat b}$ generated by the first term.
In general, one has a system of three bosonic modes, 
coupled by a bilinear interaction. The corresponding
Hamiltonian evolution of the system
can be straightforwardly obtained. This Hamiltonian description 
satisfactorily reproduces the dynamics as long as the dissipative
coupling of the mirror vibrational mode with its environment is negligible.
This happens in the case of modes with a high-Q mechanical quality factor.
In this case, the mechanical frequency $\Omega$ is sufficiently high 
(some MHz) so that the RWA of Eq.~(\ref{rwa}) can be made, 
and at the same time we can consider an interaction time, i.e., a
time duration of the incident laser pulse, much smaller than the 
relaxation time of the vibrational mode (which can be of order of one
second).

The system dynamics can be easily studied through
the (normally ordered) characteristic function $\Phi(\mu,\nu,\zeta)$,
where $\mu,\nu,\zeta$ are the complex variables corresponding
to the operators ${\hat a}_{1},{\hat b},{\hat a}_{2}$ respectively.
From the Hamiltonian (\ref{eq:Heff}) the dynamical equation for 
$\Phi$ results
\begin{eqnarray}\label{eq:Phidot}
    {\dot\Phi}&=&\chi\left(
    \mu\nu+\mu^{*}\nu^{*}-\mu^{*}\frac{\partial}{\partial\nu}
    -\mu\frac{\partial}{\partial\nu^{*}}
    -\nu^{*}\frac{\partial}{\partial\mu}
    -\nu\frac{\partial}{\partial\mu^{*}}\right)\Phi
    \nonumber\\
    &&+\theta\left(
    \zeta^{*}\frac{\partial}{\partial\nu^{*}}
    +\zeta\frac{\partial}{\partial\nu}
    -\nu^{*}\frac{\partial}{\partial\zeta^{*}}
    -\nu\frac{\partial}{\partial\zeta}\right)\Phi\,,
\end{eqnarray}
with the initial condition
\begin{equation}\label{eq:Phiini}
    \Phi(t=0)=\exp\left[-\overline{n}|\nu|^{2}\right]\,,
\end{equation}
corresponding to the vacuum for modes ${\hat a}_{1}$,
${\hat a}_{2}$ and to a thermal state for mode ${\hat b}$.
The latter is characterized by an average number of excitations
$\overline{n}=[\exp(\hbar\Omega/k_{B}T)-1]^{-1}$,
$T$ being the equilibrium temperature and $k_{B}$ the
Boltzmann constant.
Since the initial condition is a Gaussian state of the three-mode system,
and the system is linear, the joint state of the whole system at time
$t$ is still Gaussian, with characteristic function
\begin{eqnarray}\label{eq:Phisol}
    \Phi&=&\exp \left[
    -{\cal A}|\mu|^{2}-{\cal B}|\nu|^{2}-{\cal E}|\zeta|^{2} \right.
    \nonumber\\
    &&\left.
    +{\cal C}\mu\nu+{\cal C}\mu^{*}\nu^{*}
    +{\cal F}\mu\zeta+{\cal F}\mu^{*}\zeta^{*}
    +{\cal D}\nu\zeta^{*}+{\cal D}\nu^{*}\zeta\right]\,,
\end{eqnarray}
where
\begin{eqnarray}
    {\cal A}(t)&=&\frac{1}{2(r^2-1)^2}\left\{
    \left[1-\cos\left(2\Theta t\right)\right]\left[\overline{n}(r^2-1)-1\right]
    \right.\nonumber\\
    &&\left.
    +4r^2 \left[1-\cos\left(\Theta t\right)\right]\right\}\,,
    \label{eq:A}
    \\
    {\cal B}(t)&=&\frac{1}{2(r^2-1)}\left\{
    \left[1+\cos\left(2\Theta t\right)\right]\overline{n}(r^2-1)
    +1-\cos\left(2\Theta t\right)\right\}\,,
    \label{eq:B}
    \\
    {\cal C}(t)&=&\frac{1}{2(r^2-1)^{3/2}}\left\{
    2r^2
    \sin\left(\Theta t\right)\right.
    \nonumber\\
    &&\left.
    +\left[\overline{n}(r^2-1)-1\right]
    \sin\left(2\Theta t\right)\right\}\,,
    \label{eq:C}
    \\
    {\cal D}(t)&=&\frac{-r}{2(r^2-1)^{3/2}}\left\{
    \sin\left(\Theta t\right)
    +\left[\overline{n}(r^2-1)-1\right]
    \sin\left(2\Theta t\right)\right\}\,,
    \label{eq:D}
    \\
    {\cal E}(t)&=&\frac{r^2}{2(r^2-1)^2}\left\{
    \left[1-\cos\left(2\Theta t\right)\right]
    \left[\overline{n}(r^2-1)-1\right]\right.
    \nonumber\\
    &&\left.
    +4\left[1-\cos\left(\Theta t\right)\right]\right\}\,,
    \label{eq:E}
    \\
    {\cal F}(t)&=&\frac{r}{2(r^2-1)^2}\left\{
    \left[1-\cos\left(2\Theta t\right)\right]
    \left[\overline{n}(r^2-1)-1\right]\right.
    \nonumber\\
    &&\left.
    +2(1+r^2) \left[1-\cos\left(\Theta t\right)\right]\right\}\,,
    \label{eq:F}
\end{eqnarray}
$r=\theta/\chi=\left[(\omega_0+\Omega)/(\omega_0-\Omega)\right]^{1/2}$
and $\Theta=\chi (r^2-1)^{1/2}$.
The corresponding density operator 
can be expressed as
\begin{eqnarray}\label{eq:rho1b2}
    {\hat\rho}_{1b2}=\int\frac{d^{2}\mu}{\pi}
    \int\frac{d^{2}\nu}{\pi}\int\frac{d^{2}\zeta}{\pi}
    &&\Phi(\mu,\nu,\zeta,t)e^{-|\mu|^{2}-|\nu|^{2}-|\zeta|^{2}}
    \nonumber\\
    &&\times{\hat D}_{1}(-\mu){\hat D}_{b}(-\nu){\hat D}_{2}(-\zeta)\,,
\end{eqnarray}
where ${\hat D}_i$ indicates the corresponding
normally ordered displacement operator \cite{GLAUBER}.

Here we are interested in the reduced state of the
system composed by the two reflected optical sideband modes. They
do not interact directly but their interaction is mediated by the 
optomechanical coupling of each mode with the moving mirror.
This reduced state can be immediately obtained by tracing over the mirror
mode ${\hat b}$, obtaining
\begin{eqnarray}\label{eq:Phitil}
    \Phi_{12}(\mu,\zeta) 
    =\exp\left[-{\cal A}
    |\mu|^{2}
    -{\cal E}
    |\zeta|^{2}+{\cal F}\left(\mu\zeta+\mu^{*}\zeta^{*}\right)
    \right]\,.
\end{eqnarray}
Introducing the vector of field quadratures
${\hat {\bf v}}=({\hat X}_{1},{\hat P}_{1},{\hat 
X}_{2},{\hat P}_{2})$, where
\begin{equation}
    {\hat X}_{j}=\frac{{\hat a}_{j}+{\hat a}_{j}^{\dag}}{\sqrt{2}}\,,
    \quad
    {\hat P}_{j}=\frac{{\hat a}_{j}
    -{\hat a}_{j}^{\dag}}{i\sqrt{2}}\,,
    \quad {j=1,2}\;,
\end{equation}
it is possible to connect the characteristic function (\ref{eq:Phitil})
with the correlation matrix $\Gamma$ of the two sideband
modes, defined as $\Gamma_{i,j}=
\langle{\hat{\bf v}}_{i}{\hat{\bf v}}_{j}
+{\hat{\bf v}}_{j}{\hat{\bf v}}_{i}\rangle/2
-\langle{\hat{\bf v}}_{i}\rangle\langle{\hat{\bf v}}_{j}\rangle$,
and which, in this case, is equal to
\begin{equation}
    \Gamma=\left(
    \begin{array}{cccc}
    {\cal A}+\frac{1}{2}&0
    &{\cal F}&0
    \\
    0&{\cal A}+\frac{1}{2}
    &0&-{\cal F}
    \\
    {\cal F}&0
    &{\cal E}+\frac{1}{2}&0
    \\
    0&-{\cal F} &0&{\cal E}+\frac{1}{2}
    \end{array}
    \right)\,.
    \label{eq:Gam}
\end{equation}
The Gaussian state of the two optical modes is completely characterized by
this correlation matrix. In particular we can employ the necessary and 
sufficient criterion for entanglement in the case of Gaussian states
derived by Simon \cite{SIM}, which allows us to determine the parameter region
where the two modes are entangled by the optomechanical interaction with the 
mirror vibrational mode. Simon's necessary and sufficient criterion 
for entanglement becomes, for the two-sideband modes state 
considered here,
\begin{eqnarray}
{\cal T}&\equiv&\left({\cal A}{\cal E}+\frac{1}{4}+\frac{{\cal A}
+ {\cal E}}{2}-
{\cal F}^2\right)^2 +\frac{1}{16}-\frac{{\cal F}^2}{2} \nonumber \\
&&-\left(\frac{{\cal A}}{2}+\frac{1}{4}\right)^2-
\left(\frac{{\cal E}}{2}+\frac{1}{4}\right)^2 < 0 \label{simon}
\end{eqnarray}
As it can be immediately seen from 
Eqs.~(\ref{eq:A})-(\ref{eq:F}), 
the dynamics of the system is determined only by
three dimensionless parameters, $\Theta t$, $r$, and $\overline{n}$, and
it is periodic in $\Theta t$ with period $2 \pi$.
The parameter $r$ is determined by the ratio between the mechanical and
the driving laser frequency, $\Omega/\omega_0$ and it is always very close
to $1$. For this reason in Fig.~2 we show the marker of entanglement 
${\cal T}$ as a function of the scaled time $\Theta t$ for
$0\le \Theta t \le 2\pi$, for
different values of the initial mean thermal vibrational number 
$\overline{n}$ and at a fixed value of $r$, i.e.,
$r-1=2.5 \times 10^{-7}$.
Fig.~2 clearly shows that the two sideband modes are almost always entangled
(obviously except for $t=2 k \pi$, for integer $k$, due to the 
factorized initial condition) and that this entanglement is extremely robust
with respect to the thermal noise acting on the mirror.
In fact, the marker of entanglement ${\cal T}$ remains always negative,
even at extremely large values of $\overline{n}$. Fig.~2 shows 
${\cal T}$ for $\overline{n}=0$ (full line), $\overline{n}=10^5$ (dashed line,
almost coinciding with that for $\overline{n}=0$), 
$\overline{n}=5 \times 10^6$ (dotted-dashed line), and 
$\overline{n}= 10^7$ (dotted line). 
This result is extremely interesting because it shows that 
radiation pressure proves to be an efficient source of entangled
travelling optical fields, not particularly affected by thermal noise.
In this treatment we have only considered this kind of noise and we have 
neglected other more technical noise sources, such as the fluctuations
of the intensity and the frequency of the incident driving field.
In fact, the latter, differently from thermal noise, 
are negligible in today experiments involving 
optomechanical systems \cite{EXP}.

\begin{figure}
\begin{center}
\includegraphics[width=0.8\textwidth]{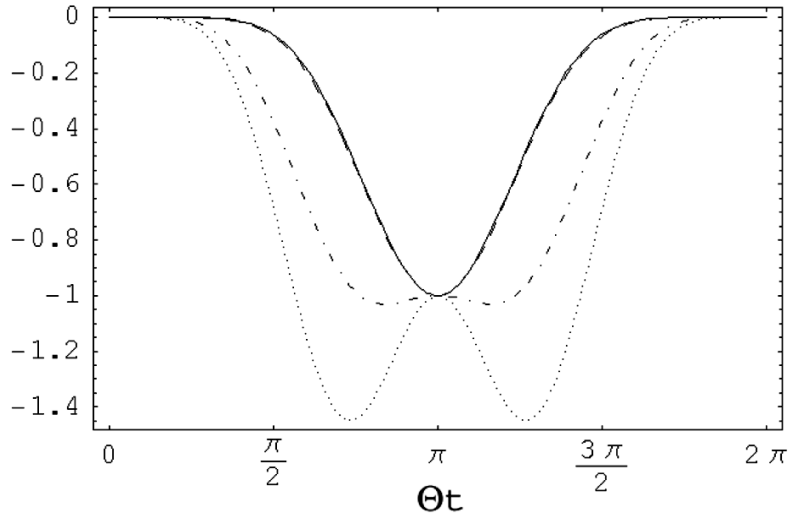}
\end{center}
\vspace{-0.5cm} \caption{\label{fig2} 
The marker of entanglement ${\cal T}(\Theta t)/|{\cal T}(\pi)|$ 
vs the scaled time 
$\Theta t$. The values of other parameters are:
$r=1+2.5\times 10^{-7}$ and
$\overline{n}=0$ solid line,
$\overline{n}=10^{5}$ dashed line,
$\overline{n}=5\times 10^{6}$ dashed-dotted line,
$\overline{n}=10^{7}$ dotted line.
}
\end{figure}

\begin{figure}
\begin{center}
\includegraphics[width=0.5\textwidth]{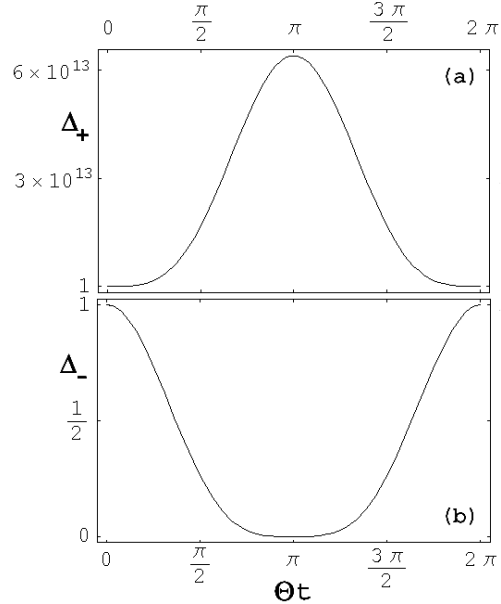}
\end{center}
\vspace{-0.5cm} \caption{\label{fig3} 
The quantities $\Delta_{+}(\Theta t)$
(a) and 
$\Delta_{-}(\Theta t)$
(b) are plotted vs the scaled time 
$\Theta t$. The values of other parameters are
$r=1+2.5\times 10^{-7}$ and
$\overline{n}=10^{5}$.

}
\end{figure}

From Fig.~2, it is evident that
${\cal T}$ at exactly half period, $\Theta t = \pi$, 
is unaffected by thermal noise, i.e., it is independent of $\overline{n}$.
In fact, using Eqs~(\ref{eq:A})-(\ref{eq:F}) and (\ref{simon}), 
it is easy to see that
${\cal T}(\Theta t=\pi)= -4(r+r^3)^2/(r^2-1)^4$ 
(actually ${\cal T}$ has been scaled exactly 
by the modulus of this value in Fig.~2). 
This independence of $\overline{n}$ is a 
manifestation of quantum interference between the dynamical effects 
of the two interaction terms in Eq.~(\ref{eq:Heff}).
It is therefore interesting to see the reduced state of the two reflected
sideband modes just at this interaction time, $\Theta t = \pi$.
Using Eqs.~(\ref{eq:A})-(\ref{eq:F}), one has the following correlation matrix
\begin{eqnarray}
    &&\Gamma(\Theta t = \pi)=\\
    &&=\left(
    \begin{array}{cccc}
    \frac{4r^2}{(r^2-1)^2}+\frac{1}{2}&0
    &\frac{2r(1+r^2)}{(r^2-1)^2}&0
    \\
    0&\frac{4r^2}{(r^2-1)^2}+\frac{1}{2}
    &0&-\frac{2r(1+r^2)}{(r^2-1)^2}
    \\
    \frac{2r(1+r^2)}{(r^2-1)^2}&0
    &\frac{4r^2}{(r^2-1)^2}+\frac{1}{2}&0
    \\
    0&-\frac{2r(1+r^2)}{(r^2-1)^2} &0&\frac{4r^2}{(r^2-1)^2}+\frac{1}{2}
    \end{array}
    \right)\,.\nonumber
    \label{eq:Gam2}
\end{eqnarray}
It is easy to check that this is just the correlation matrix of an EPR-like
entangled state, i.e., identical to the two-mode squeezed state generated by
a parametric amplifier \cite{QO94}. 
The correspondence between the squeezing parameter
$\xi$ and our parameters is
$\sinh \xi \leftrightarrow 2r(1+r^2)/(r^2-1)^2$, so that, in practice 
one can reach very high two-mode squeezing by choosing $r-1 \rightarrow 0$,
i.e., by decreasing the ratio $\Omega/\omega_0$.
This fact can be equivalently seen by evaluating the variances of the
linear combination of field quadratures, 
typical of the two-mode squeezed state, i.e.,
\begin{eqnarray}
\Delta_{-}&=&\langle (X_1-X_2)^2\rangle = \langle (P_1+P_2)^2\rangle 
\,,\label{eq:D-} \\
\Delta_{+}&=&\langle (X_1+X_2)^2\rangle = \langle (P_1-P_2)^2\rangle  
\label{eq:D+}\,.
\end{eqnarray}
For $\Theta t=\pi$, one has
\begin{eqnarray}
\Delta_{-}(\Theta t=\pi)&=&\left(\frac{r-1}{r+1}
\right)^2\,, \\
\Delta_{+}(\Theta t=\pi)&=&
\left(\frac{r+1}{r-1}
\right)^2 \,.
\end{eqnarray}
The time behavior of the variances (\ref{eq:D-}) and (\ref{eq:D+})
is depicted in Fig.~\ref{fig3}.
It is worth noticing that the typical EPR correlations
are available for a wide time range around $\Theta t=\pi$.
Furthermore, they are robust against thermal noise since 
the curves are shown for ${\overline n}=10^5$, and they practically 
coincide with those for ${\overline n}=0$.
That is in agreement with the results shown in Fig.\ref{fig2}.

\section{Results and Conclusions}

In conclusion, we have seen that, under appropriate
conditions, the radiation pressure of an intense optical pulse
impinging on a perfectly reflecting mirror couples efficiently
the first two sideband modes induced by a vibrational mode
of the mirror. This optomechanical interaction actually
entangles the two modes, and quite surprisingly, the resulting
entanglement is extremely robust with respect to the thermal noise 
acting on the mirror. In particular, due to quantum interference,
if the interaction time, i.e., the duration of the incident driving
pulse, is appropriately tailored, the reduced state of the 
reflected sideband modes is a two-mode squeezed state.
Therefore, rather unexpectedly,
radiation pressure could be employed as a new source
of entangled two-mode squeezed state, in which the
role of the squeezing parameter is played by the ratio between the
optical and the mechanical frequency $\omega_0/\Omega$.
These interesting results hold if the time duration of the pulse satisfies
two conditions: i) it has to be much longer than the period of mechanical
oscillations, because we have to average over these oscillations;
ii) it has to be shorter than the vibrational mode relaxation time
(mechanical damping has been neglected in our treatment).
These conditions are not extremely easy but they could be realized
using micro-opto-electro-mechanical-systems (MOEMS) for example
\cite{MOEMS}.
Possible values are: damping rates $\gamma_{m} \simeq 1$ Hz,
${\wp} = 10$W, $\omega_{0}\sim 2\times 10^{15}$ Hz, $\Omega \sim 
5\times 10^{8}$ Hz, $\Delta\nu_{det}\sim 10^{7}$ Hz, 
$\Delta\nu_{mode}\sim t^{-1} \sim 10^{3}$ Hz, and $M \sim 10^{-10}$ 
Kg, yielding
$\chi \simeq \theta \simeq 5\times 10^{5}$ Hz, and $\Theta \simeq 
10^{3}$ Hz. Therefore, using well tailored and long pulses (order
of $1$ ms), one could realize this new source of two mode squeezing.

\end{document}